\documentstyle[12pt,rotating,epsfig,glas]{article}
\begin{document}
\def\AP#1#2#3{Ann.\ Phys.\ (NY) #1 (19#3) #2}
\def\CPC#1#2#3{Computer Phys. Comm. #1 (19#3) 32}
\def\PL#1#2#3{Phys.\ Lett.\ #1B (19#3) #2}
\def\PR#1#2#3{Phys.\ Rev.\ #1D (19#3) #2}
\def\NP#1#2#3{Nucl.\ Phys.\ #1B (19#3) #2}
\def\NC#1#2#3{Nuovo Cimento\ #1A (19#3) #2}
\def\ZP#1#2#3{Z.\ Phys.\ C#1 (19#3) #2}
\def\JL#1#2#3{JETP Lett.#1 (19#3) #2}
\def\JP#1#2#3{J.\ Phys.\ G#1 (19#3) #2}
\def\NIM#1#2#3{Nucl.\ Instr.\ Meth. A#1 (19#3) #2}
\def\RMP#1#2#3{Rev.\ Mod.\ Phys. #1 (19#3) #2}
\def\z{z}
\def\gsubh{\gamma_{H}}
\def\qsd{Q^2}
\def\mjj{\protect\sqrt{\hat{s}}}
\def\mj2{\protect\hat{s}}
\newcommand{\logxp}     {\mbox{$\xi=\ln(1/x_{p})$}}
\newcommand{\logxpmax}  {\mbox{$\xi_{\rm peak}$}}
\newcommand{\as}{\mbox{$\alpha_{s}$}}
\newcommand{\mz}{\mbox{$m_{Z^0}$}}
\newcommand{\xp}{\mbox{$x_{p}$}}
\newcommand{\ee}{\mbox{$e^+e^-$}}
\newcommand{\qqbar}{\mbox{$q{\overline q}$}}
\newcommand{\ppbar}{\mbox{$p{\overline p}$}}
\newcommand{\ycut}{\mbox{$y_{cut}$}}
\newcommand{\GeV}{\mbox{GeV}}
\newcommand{\als}{$\alpha_s$}
\newcommand{\ep}{\mbox{$e^{\pm}p$}}
\newcommand{\mup}{\mbox{$\mu^{\pm}p$}}
\newcommand{\pp}{\mbox{$p\bar{p}$}}
\def\lms{\Lambda_{\overline{MS}}}
\def\mz{M_{Z^0}}

\begin{titlepage}{GLAS-PPE/97--08}{October 1997}
\title{Hadronic Final States in Deep Inelastic Scattering
at HERA }
\author{N.~H.~Brook}
\conference{Lecture given at ``The Actual Problems of Particle
Physics'', Gomel, Belarus.}
\begin{abstract}
This lecture contains a brief introduction to HERA and 
deep inelastic scattering (DIS),
before going on to highlight some of the
 measurements of the hadronic final state in DIS
performed by the H1 and ZEUS collaborations.
\end{abstract}
\end{titlepage}
\section{The HERA Accelerator and Detectors}
The HERA accelerator, located at DESY in Hamburg, is an
electron-proton collider. It is $6.3~{\rm\ km}$ in circumference and
collides positrons (or electrons) at $\approx 27{\rm \ GeV}$
with protons at $820{\rm \ GeV}.$ There are four interaction 
regions:
two containing general purpose, 
hermetic detectors (H1 and ZEUS); another experiment
(HERMES) investigating the spin distributions of the quarks in protons
and neutrons; and another (HERA-B) planning to measure CP violation in the
B-system. The H1 and ZEUS detectors took first data in 1992.

\begin{figure}[htb]
\vspace*{-1.0cm}
\mbox{\epsfig{file=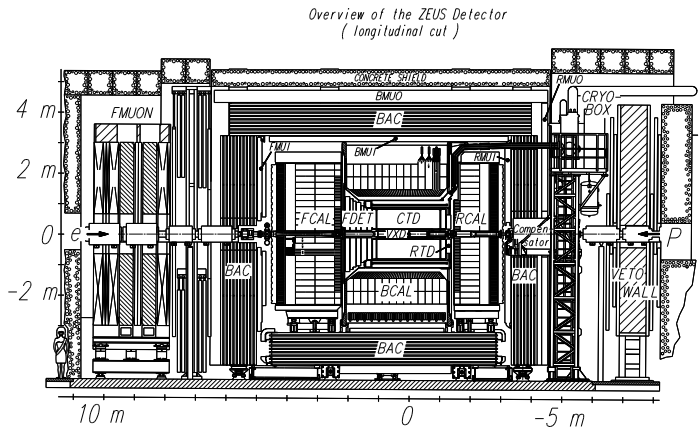,height=8.0cm,angle=270.}}
\vspace*{-3.5cm}
\caption{\it Cross sectional view of the ZEUS detector}
\label{fig:ZEUS}
\end{figure}

The ZEUS detector is shown in figure~\ref{fig:ZEUS}. The asymmetric
design of the detector reflects the proton energy being significantly
higher than that of the electron beam. 

The tracking system consists of a vertex detector (VXD)~\cite{b:VXD}
and a central tracking chamber (CTD)~\cite{b:CTD}
enclosed in a 1.43 T solenoidal magnetic field.
Immediately surrounding the beampipe is the VXD  which
consists of 120 radial cells, each with 12 sense wires.
The CTD, which encloses the VXD,  is a drift chamber consisting of
72~cylindrical layers, arranged in 9 superlayers. Superlayers with
wires parallel to the beam axis alternate with those inclined at a small
angle to give a stereo view.
A forward tracking detector is employed in the forward region to detect
tracks in the proton direction and consists of three 12-layer planar drift
chambers sandwiched with pairs of transition radiation detectors. In the
rear direction there
is an additional 12-layer planar drift chamber known as the rear
tracking detector (RTD).

Outside the solenoid is the uranium-scintillator
calorimeter (CAL)~\cite{b:CAL},
which is divided into three parts:
forward, barrel and rear covering the polar regions
$2.6^\circ$ to $36.7^\circ$,
$36.7^\circ$ to $129.1^\circ$ and
$129.1^\circ$ to $176.2^\circ$, respectively.
The CAL covers 99.7$\%$ of the solid angle, with
holes of $ 20 \times 20 $ cm$^{2}$ in the centres of
the forward and rear calorimeters to
accommodate the HERA beam pipe. Each of the calorimeter parts is
subdivided
into towers which are segmented longitudinally into electromagnetic
(EMC)
and hadronic (HAC) sections. 
The small angle rear tracking
detector (SRTD)~\cite{z_shift}, which is attached to
the front face of the rear calorimeter, measures the impact point of
charged
particles at small angles with respect to the positron beam direction.

The iron return yoke for the magnet is instrumented with
proportional counters. This backing calorimeter (BAC) measures any hadronic
energy which `leaks out' out of the main calorimeter. Beyond that and
in the forward direction there are further detectors for muon detection.

Downstream of the main detector in the proton direction, six measuring
stations are installed in the proton ring for detecting forward scattered
protons. Beyond the final station, further downstream, is a forward
neutron calorimeter.
In the electron direction, two lead scintillator calorimeters placed
$-35{\rm\ m}$ 
and $-107{\rm\ m}$ from the interaction point measure the luminosity
and tag events with a small momentum transfer~\cite{b:LUMI}.

A fuller
description of the ZEUS detector can be found in reference~\cite{b:Detector}.
The H1 detector is of a very similar layout as ZEUS and a description can
be found in reference~\cite{H1app}.

\section{DIS Kinematics}
The event kinematics of deep inelastic scattering, DIS,
are determined by the negative square of the four-momentum transfer at
the positron vertex,
$Q^2\equiv-q^2$, and the Bjorken scaling variable, $x=Q^2/2P \cdot q$,
where $P$ is the four-momentum of the proton.
In the Quark Parton Model (QPM),
the interacting quark from the proton carries the four-momentum $xP$.
The variable $y$, the fractional energy transfer to the proton in its
rest
frame, is related to $x$ and $Q^2$ by $y\simeq Q^2/xs$, where $\sqrt s$
is
the positron-proton centre of mass energy.
Because the H1 and ZEUS detectors are almost hermetic the kinematic
variables $x$ and $Q^2$ can be reconstructed in a variety of 
ways using combinations of electron and hadronic system energies and
angles~\cite{DA}.

\begin{figure}[htb]
\vspace*{-1.0cm}
\centerline{\epsfig{file=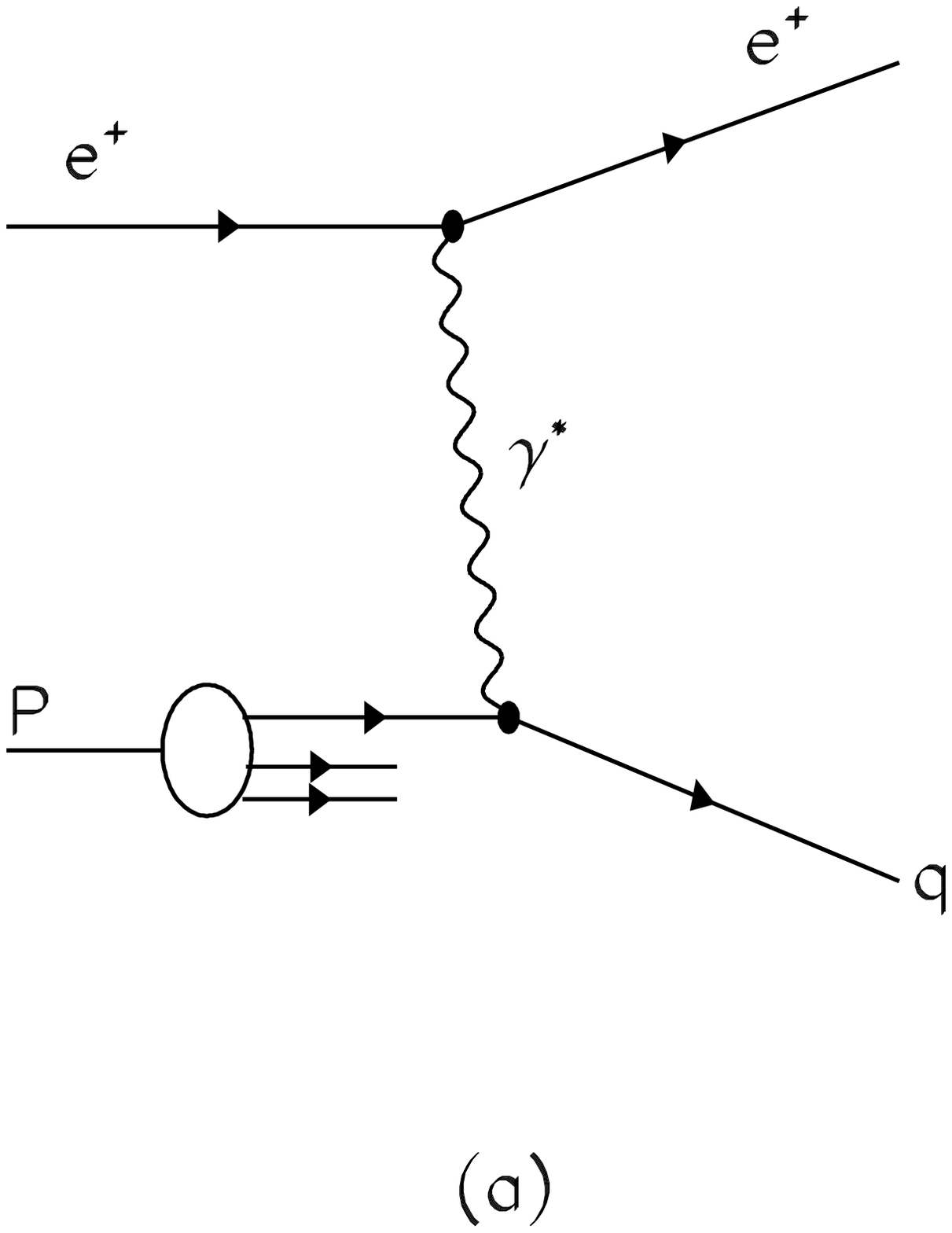,width=4.4cm}
\epsfig{file=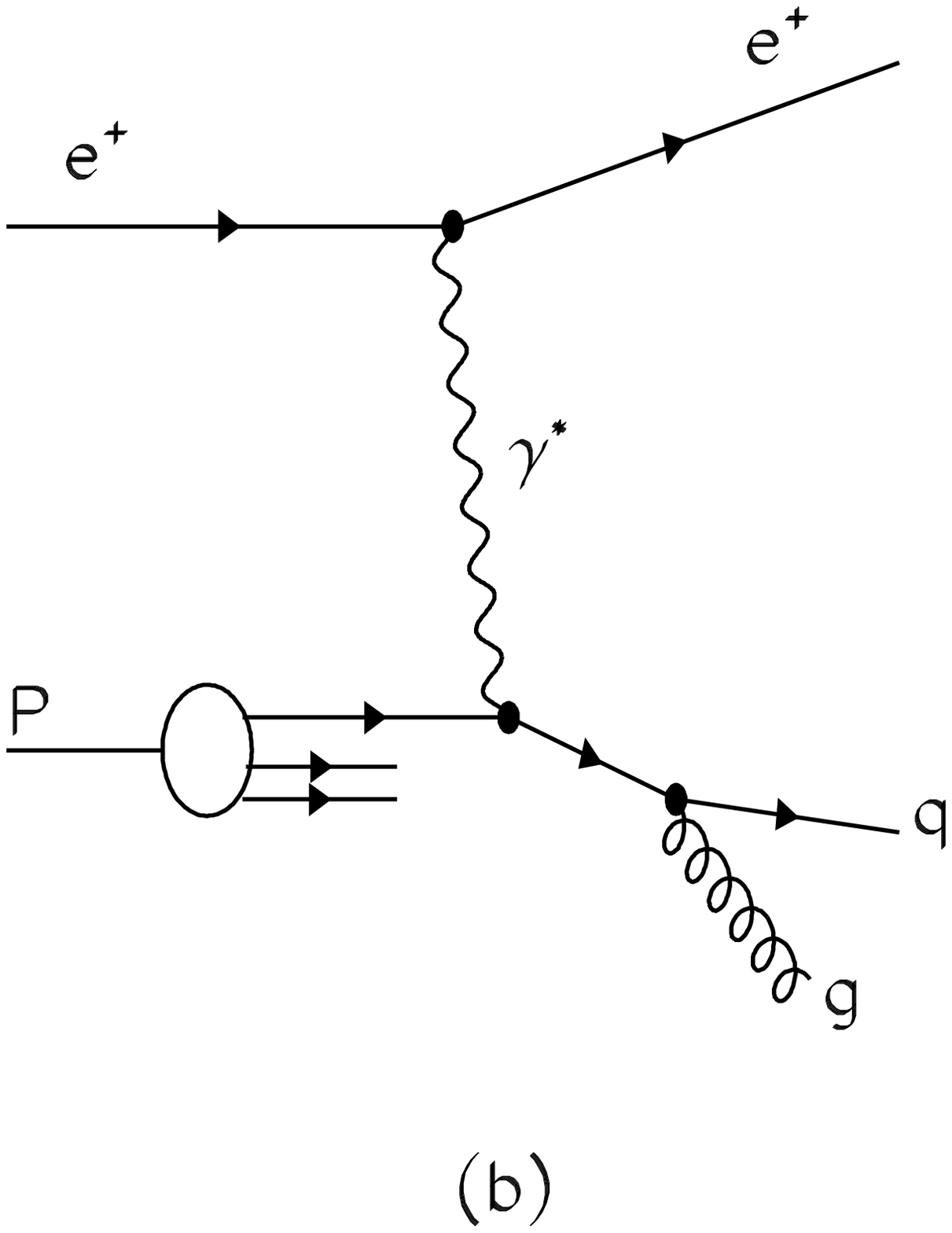,width=4.4cm}
\epsfig{file=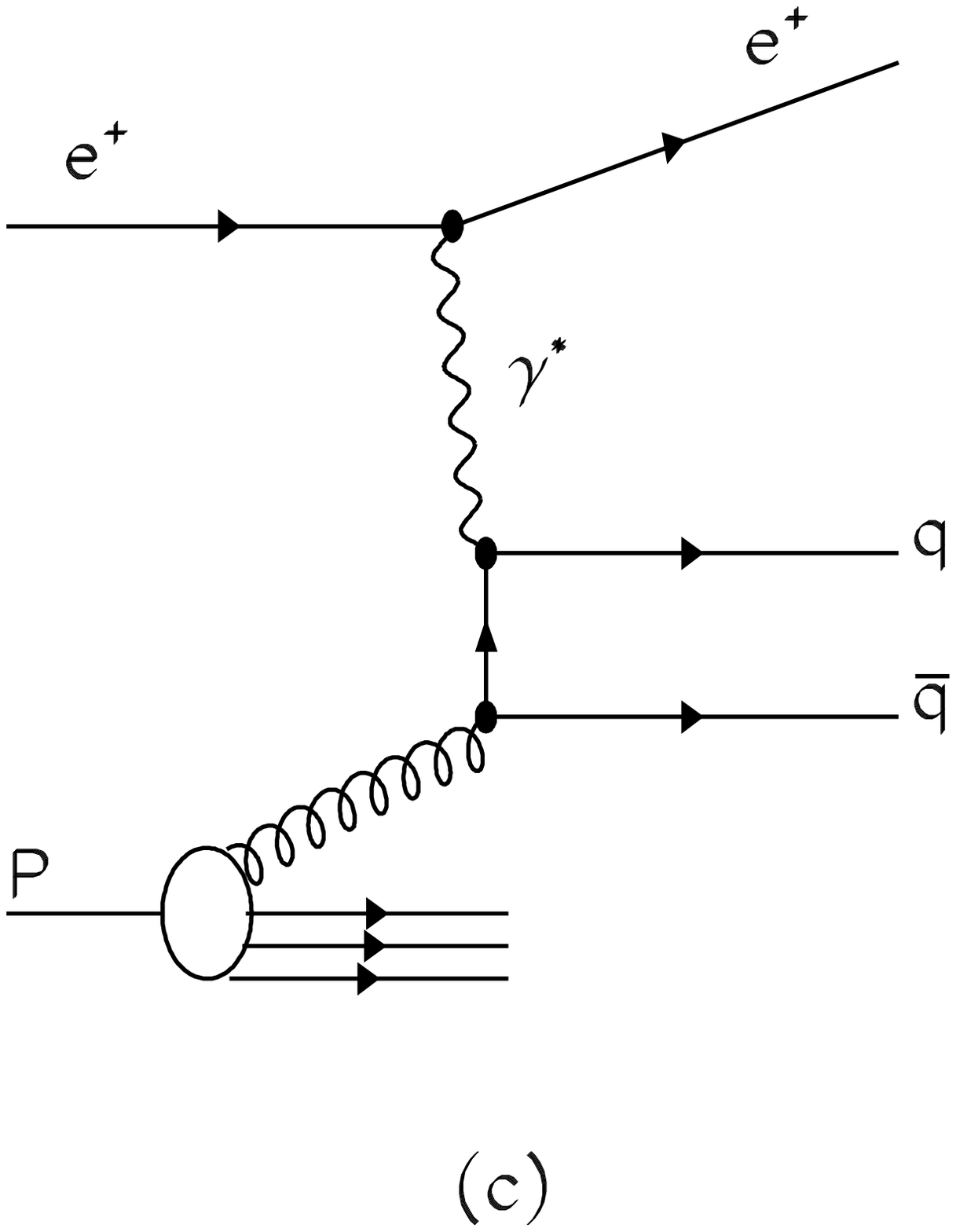,width=4.4cm}}
\caption{(a) QPM (b) QCDC  and (c) BGF diagrams}
\label{fig:feyn_alphas}
\end{figure}

In QPM
there is a
1+1 parton configuration, fig.~\ref{fig:feyn_alphas}a,
which consists of a single 
struck quark and the proton remnant,
denoted by ``+1''.
At HERA energies there are significant higher-order 
Quantum Chromodynamic (QCD) corrections:
to leading order in the strong coupling constant, $\alpha_{\rm s},$
these are QCD-Compton scattering (QCDC), 
where a gluon is radiated by the
scattered quark and Boson-Gluon-Fusion (BGF),
where the virtual boson and a gluon fuse to
form a quark-antiquark pair.
Both processes have 2+1 partons in the final
state, as shown in fig.~\ref{fig:feyn_alphas}. 
There also exists calculations for the higher, 
next-to-leading (NLO) processes.

Perturbative QCD does not predict the absolute value of the parton
densities within the proton but determines how they vary from a
given input. For a given initial distribution at a particular scale
Altarelli-Parisi (DGLAP) evolution~\cite{dglap} enables the
distributions at higher $Q^2$ to be determined.
DGLAP evolution resums the leading 
$\log(Q^2)$ contributions associated with a chain of gluon emissions.
At large enough electron-proton centre-of-mass energies there is a second
large variable $1/x$ and, therefore, it is also necessary 
to resum the $\log(1/x)$ contributions.
This is acheived by using the BFKL equation~\cite{bfkl}.

\section{Jet Physics}
To relate the hadronic final state to
the underlying hard partonic behaviour
it is generally necessary to apply a jet algorithm.
The JADE algorithm~\cite{JADE} has been used in the following analyses
as it was, at the time, the only algorithm
which allowed comparison to the NLO calculations (PROJET~\cite{PROJET}
 and DISJET~\cite{DISJET}).
The JADE algorithm
is a cluster
algorithm based on
the scaled invariant mass-squared
$$y_{ij}^{\rm JADE}=
\frac{2 E_i E_j(1-\cos\theta_{ij})}{W^2}$$
for any two objects $i$ and $j$ 
assuming that these objects are massless.
$W^2$ is the squared invariant mass
of the hadronic final state and $\theta_{ij}$ is the angle
between the two objects of energies $E_i$ and $E_j$.
The minimum $y_{ij}$
of all possible combinations is found. If
the value  of this minimum $y_{ij}$  is less than
the variable cut-off parameter $\ycut$, the two objects $i$ and
$j$ are
merged into a new object by adding their four-momenta and the process is
repeated until all $y_{ij}>\ycut$. The surviving
objects are called jets which represent the underlying partonic
structure that is dependent on \als.

\begin{figure}[htb]
\centerline{ 
\epsfig{file=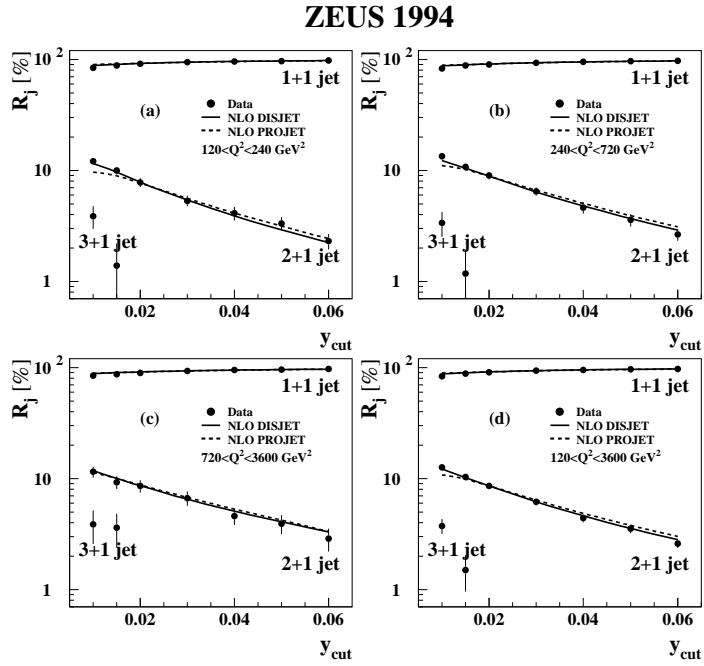,height=10.0cm}}
\vspace*{-0.7cm}
\caption{ \it Jet production rates $R_j$ as a function of the jet
          resolution parameter \ycut\ for $Q^2$ in the range
          (a) $120<Q^2<240$~GeV$^2$, (b) $240<Q^2<720$~GeV$^2$,
          (c) $720<Q^2<3600$~GeV$^2$, and (d) $120<Q^2<3600$~GeV$^2$.
          Only statistical errors are shown.
          Two NLO QCD calculations, DISJET and PROJET,
          each with the value of $\lms$ obtained from the fit at
          \ycut=0.02, are also shown.
        }
\label{fig:jet_rates}
\end{figure}

Figures~\ref{fig:jet_rates}a--d show the ZEUS jet rates using data taken
in 1994,
$R_{1+1}, R_{2+1}$ 
  and $R_{3+1}$ as a function of
\ycut\ for data compared with the DISJET and PROJET NLO QCD calculations
for three $Q^2$ intervals 
$120<Q^2<240$~GeV$^2$, $240<Q^2<720$~GeV$^2$,
$720<Q^2<3600$~GeV$^2$, and the combined region
$120<Q^2<3600$~GeV$^2$.
There is good agreement between the
corrected $1+1$ and $2+1$
jet rates and the NLO QCD calculation over most of the
range in \ycut\ shown.
Both programs agree well in their prediction of the jet-rate dependence
as a function of \ycut.

\begin{figure}
\begin{center}
 \epsfig{file=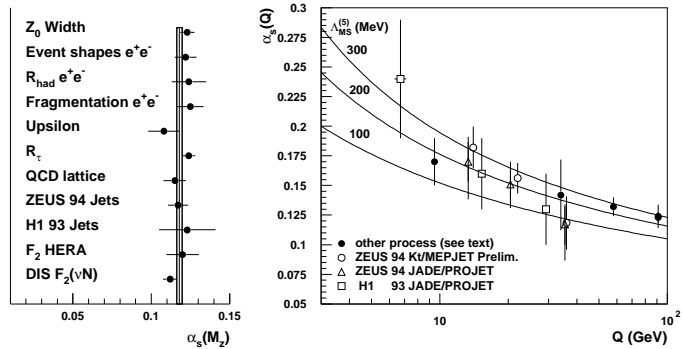,width=10.0cm,bbllx=5pt,bblly=1pt,bburx=604pt,bbury=290pt}
\end{center}
\vspace*{-0.5cm}
\caption{\it Left: Values and total error of $\alpha_s(M_Z)$ from
 various processes.
 The solid line indicates the world average 
 and the band its total error.
 Right: \als($Q$) from HERA (open symbols) and other processes
 with increasing $Q$ (closed circles):
 $\Gamma_{\! \Upsilon}$ and
 $\sigma_{\rm \! had}/\sigma_{\rm \! tot}$,
 event shapes and $\Gamma_{\rm \! hadron}/\Gamma_{\rm \! lepton}$
 in \ee.
\label{fig:alphas}}
\end{figure}

The values of \als($Q$) extracted
by the H1~\cite{h1alphas} and ZEUS~\cite{zeusalphas}
collaboration as a function of $Q$ 
are shown in Fig.~\ref{fig:alphas}. 
The value of \as\ was determined by varying the $\Lambda$ scale parameter
in the QCD
calculation until the best fit to the ratio $R_{2+1}$ was obtained
at a particular value of \ycut.
The measured \as\ decreases with increasing $Q$, consistent
with the running of the strong coupling constant, with $Q^2$ taken as
the
scale.
In addition the figure shows the curves for
$\lms^{(5)}$ = 100, 200, and 300~MeV.
An extrapolation to \als($M_z$) yields:
\begin{eqnarray}
\vspace{-0.1cm}
\label{eq:alphas}
 \nonumber \! \! \! \! \! \! \! \! \! \!
 {\rm H1 \; 93:  } \; \; \; \; \; \; \; \; \; \;
  \alpha_s(M_z)  = 0.123 \pm 0.012 {\rm(stat)}
 \pm 0.013 {\rm (syst.)} \\
 \nonumber
 {\rm ZEUS \; 94:} \; \; \alpha_s(M_z) = 0.117
 \pm 0.005 {\rm (stat)}^{ + 0.005}_{-0.004} {\rm(exp.)} \pm 0.007 {\rm (th.)}
\end{eqnarray}
which are consistent with other values obtained from
a large variety of different processes as shown Fig.~\ref{fig:alphas}
(see~\cite{pdg} for references).
Even with the current statistics
the HERA measurements are already
competitive with those made elsewhere.

Recently two new, more flexible
NLO calculations (MEPJET~\cite{mepjet} and
DISENT~\cite{disent}) have become
 available allowing the experiments
to analyze the data using any particular jet algorithm.
The $k_T$ algorithm~\cite{ktalgo} is particularly suited for DIS
as it allows factorization between the beam fragmentation and the hard
process~\cite{durweb}.
The ZEUS collaboration has reanalyzed~\cite{trefzger} their 1994 data using
this algorithm. 
The preliminary values of \als($Q$) obtained
 in the three bins of $Q$ are shown (with statistical
 errors only) in Fig.~\ref{fig:alphas} and
are consistent with the results obtained
 with the JADE algorithm.
\section{Event Shapes}

A natural frame in which to study the dynamics of the hadronic final
state
in DIS is the Breit frame~\cite{feyn}.
In this frame the exchanged
virtual boson is purely
space-like with 3-momentum ${\bf q}=(0,0,-Q)$, the incident quark
carries momentum $Q/2$ in the positive $Z$ direction,
and the outgoing struck quark carries Q/2 in the negative
$Z$ direction.  A final state particle has a 4-momentum 
$ p^B$ in this frame,
and is assigned to the current region if $p^B_Z$ is negative, and to the
target frame if $p^B_Z$ is positive.
The advantage of this
frame lies in the maximal separation of the outgoing parton from
radiation associated with the incoming parton and the proton remnant,
thus providing the optimal environment for the study of the
fragmentation of the outgoing parton.

Event shape variables have been investigated in $e^+e^-$
experiments and used to extract the strong coupling constant
$\alpha_s(M_Z)$ independent of any jet algorithm,
see {\em eg} ref.~\cite{bethke}.
H1 have recently performed a similar analysis~\cite{H1shapes}
in deep inelastic scattering 
in the current fragmentation region
of the Breit frame. 

The event shape dependence on $Q$ (or energy dependence)
can be due to the logarithmic change of the strong coupling constant
$\alpha_s(Q) \propto 1/\ln Q$, and/or
power corrections (hadronisation effects)
which are expected to behave like $1/Q$.
Recent theoretical developments suggest that $1/Q$ corrections are not
necessarily related to hadronisation, but may instead be a universal
soft
gluon phenomenon associated with the behaviour of the running coupling
at
small momentum scales~\cite{webber,dokshitzer}.

H1 have analysed a number of
infrared safe (ie independent of the number of partons produced)
event shape variables.
Their definitions are given below,
where the sums extend over all hadrons $h$
(being a calorimetric cluster in the detector or a parton in the QCD
calculations)
with four-momentum
$p^B_h = \{ E^B_h,\, {\bf p}^B_h \}$ 
The current hemisphere axis ${\bf n}  = \{0,\,0,\,-1 \}$
coincides with the virtual boson direction.
\begin{itemize}
  \item {\bf Thrust {\boldmath $T_c$}}
    \begin{eqnarray}
      T_c & = & \max \, \frac{\sum_h |\, {\bf p}^B_h\cdot {\bf n}_T \, |}
                {\sum_h |\, {\bf p}^B_h \, |}
      \quad\quad\quad\quad\quad\quad\quad\quad \
      {\bf n}_T \ \equiv \ \mbox{thrust axis} \ ,
      \quad \ \
      \nonumber
    \end{eqnarray}
  \item {\bf Thrust {\boldmath $T_z$}}
    \begin{eqnarray}
      T_z & = & \frac{\sum_h |\, {\bf p}^B_h\cdot {\bf n} \, |}
                 {\sum_h |\, {\bf p}^B_h \, |}
          \ = \ \frac{\sum_h |\, {\bf p}^B_{z\,h}\, |}
                     {\sum_h |\, {\bf p}^B_h \, |}
      \quad\quad\quad\quad \ \ \
      {\bf n} \ \equiv  \ \mbox{hemisphere axis} \ ,
      \nonumber
    \end{eqnarray}
  \item {\bf Jet Broadening {\boldmath $B_c$}}
    \begin{eqnarray}
        B_c & = & \frac{\sum_h |\, {\bf p}^B_h\times {\bf n} \, |}
                     {2\,\sum_h |\, {\bf p}^B_h \, |}
          \ = \ \frac{\sum_h |\, {\bf p}^B_{\perp\,h}\, |}
                     {2\,\sum_h |\, {\bf p}^B_h \, |}
      \quad\quad\quad\quad
      {\bf n} \ \equiv \ \mbox{hemisphere axis} \ ,
      \nonumber
    \end{eqnarray}
  \item {\bf Scaled Jet Mass {\boldmath $\rho_c$}}
    \begin{eqnarray}
      \rho_c & = & \frac{M^2}{Q^2}
      \ = \ \frac{(\, \sum_h \, p^B_h \, )^2}{Q^2} \ .
      \phantom{xxxxxxxxxxxxxxxxxxxxxxxxxxxxxxx}
      \nonumber
    \end{eqnarray}
\end{itemize}

\begin{figure}[htb]
\vspace*{-1.0cm}
\centerline{\epsfig{file=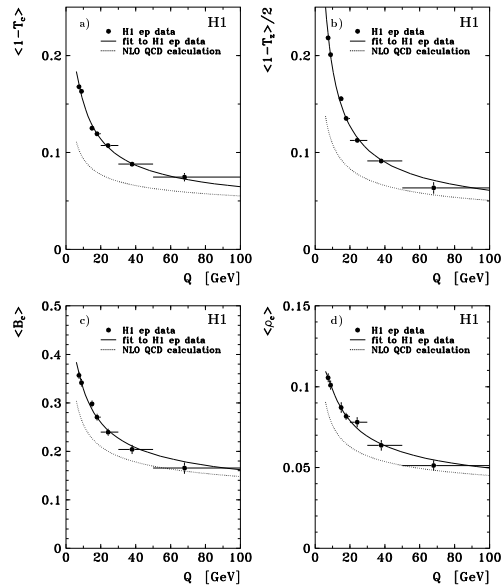,width=8.0cm}}
\vspace*{-2.0cm}
  \caption{\it Mean event shape variables as a function of Q for
           {\bf a)} $\langle 1 - T_c \rangle$,
           {\bf b)} $\langle 1 - T_z \rangle / 2$,
           {\bf c)} $\langle B_c \rangle$, and
           {\bf d)} $\langle \rho_c \rangle$.
           H1 DIS $e\,p$ data ($\bullet$, errors include statistics
           and systematics) are compared with QCD fits (---)
           and second order QCD calculations ($\cdot\cdot\cdot$)}
  \label{meandata}
\end{figure}

A common characteristic of the mean event shape values
$\langle 1 - T_c \rangle, \ \langle 1 - T_z \rangle, \langle \ B_c
\rangle$ and $\langle \rho_c \rangle$ 
is the fact that they exhibit a clear decrease with rising $Q$,
fig.~\ref{meandata}.
This is due to fact
that the energy flow becomes more collimated along the event
shape axis as $Q$ increases, a phenomenon also observed in \ee\
annihilation experiments.

H1 showed by fitting to the data in fig.~\ref{meandata} 
all the event shape variables can be well described by just the
first order power
corrections $\propto 1/Q,$ without the need for any higher order
corrections.
The second order perturbative QCD parton predictions are also shown and
their discrepancies with the data show that
the power corrections are substantial
at low
values of $Q$, but become less important with increasing energy.

The analysis of the event shapes give results consistent with each other
for $\bar{\alpha}_0,$ the power correction parameter
thus supporting the 
prediction of universality~\cite{webber},
 and also gives consistent values of $\alpha_s(M_Z).$ 
The results of the fit are
$\bar{\alpha}_0 = 0.491 \pm 0.003~({\mbox{exp}})
  \ ^{+0.079}_{-0.042}~({\mbox{theory}})$
for the power correction parameter and
$\alpha_s(M_Z) = 0.118 \pm 0.001~({\mbox{exp}})
  \ ^{+0.007}_{-0.006}~({\mbox{theory}})$
for the strong coupling constant in the $\overline{\mbox{MS}}$ scheme.
These values are compatible with those extracted by \ee\ 
experiments~\cite{eedata_shape}
\section{Fragmentation Functions}
Fragmentation functions represent the
probability
for a parton to fragment into a particular
hadron carrying a certain fraction of the parton's energy.
Fragmentation functions incorporate the long distance, non-perturbative
physics of the hadronization process in which the observed hadrons are
formed from final state partons of the hard scattering process
and, like structure functions,
cannot be calculated in perturbative QCD, but
can be evolved from a starting distribution at a defined energy scale.
If the fragmentation functions are combined with
the cross sections for the inclusive production of
each parton type in the given physical process, predictions can be made
for the scaled momentum, $x_p,$ spectra of final state hadrons.
Small $x_p$ fragmentation is significantly affected by the coherence 
(destructive interference) of
soft gluons~\cite{basics}, whilst 
scaling violation of the fragmentation function at 
large $x_p$ allows a measurement of \als~\cite{webnas}.

In \ee~annihilation the two quarks are produced
with equal and opposite momenta, $\pm \sqrt{s}/2.$
This can be compared with
a quark struck from within the
proton with outgoing momentum $-Q/2$ in the Breit frame.
In the direction of the struck quark (the current fragmentation region)
the particle momentum spectra, $x_p = 2p^B/Q,$
are expected to have a
dependence on $Q$ similar to those observed in
\ee~annihilation~\cite{eedis,anis,char} at energy $\sqrt{s}=Q.$

\begin{figure}[htb]
\vspace*{-1.0cm}
\begin{center}\mbox{\epsfig{file=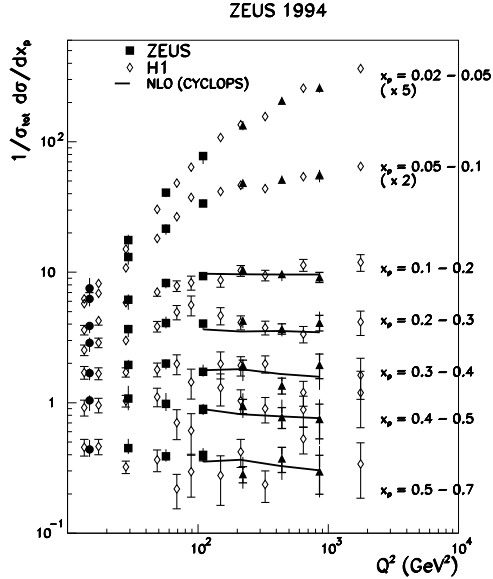,width=8cm}}
\end{center}
\vspace*{-1.0cm}
\caption{\it The inclusive charged particle distribution,
$ 1/\sigma_{tot}~ d\sigma/dx_p$,
in the current fragmentation region of the Breit frame compared to the
NLO calculation, CYCLOPS~\protect\cite{Dirk}.}
\label{NLO}
\end{figure}

The inclusive charged particle distributions~\cite{H1frag,Zfrag}, 
$(1/\sigma_{tot})d\sigma/d x_p,$ 
are shown in figure~\ref{NLO} 
plotted in bins of
fixed $x_p$ as a function of $Q^2.$
For $Q^2 > 80{\rm\ GeV^2}$ the distributions rise
with $Q^2$ at low $x_p$ and
fall-off at high $x_p$ and high $Q^2$.
By measuring the amount of
scaling violation one can ultimately
measure the amount of parton radiation and
thus determine $\alpha_s.$
Below $Q^2=80{\rm\ GeV^2}$ the fall off is due to depopulation of the
current region.

The results can be compared to
the next-to-leading order (NLO)
QCD calculations, as implemented in CYCLOPS~\cite{Dirk}, of the charged
particle inclusive distributions
in the restricted region
$Q^2 > 80 {\rm \ GeV^2}$  and
$ x_p > 0.1 ,$
where the theoretical uncertainties are small, unaffected by the 
hadron mass effects which are not included in the fragmentation function. 
This comparison is shown in figure~\ref{NLO}.
The NLO calculation combines a full next-to-leading order matrix element
with the
${\rm MRSA^{\prime}}$ parton densities (with a $\Lambda_{\rm QCD} =
230{\rm \ MeV})$
and NLO fragmentation functions
derived by
Binnewies et al. from fits to $e^+e^-$ data \cite{binnewies}.
The  data and the NLO calculations are in good agreement,
supporting the idea of universality of quark fragmentation.

\begin{figure}[hbt]
\vspace*{-1.0cm}
\centerline{\epsfig{figure=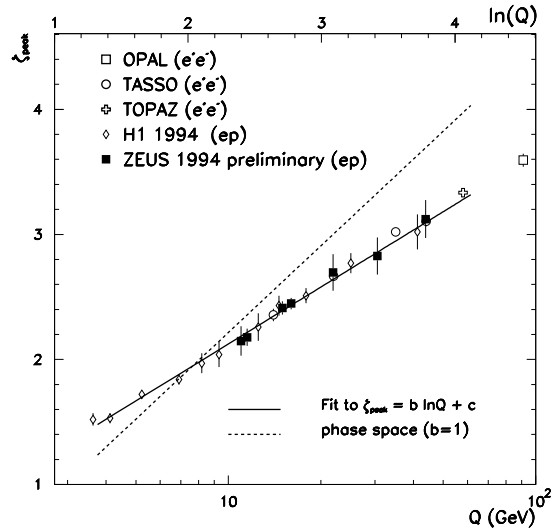,width=8cm}}
\vspace*{-0.25cm}
\caption{{\it \logxpmax~as a function of $Q.$
The HERA data are compared
to results from OPAL, TASSO and TOPAZ.
A straight line fit of the form $\logxpmax~=~b~ln(Q)~+~c$
to the ZEUS \logxpmax~values is indicated as well as the line
corresponding to $b~=~1$, discussed in the text.}}
\label{figure:slope}
\end{figure}

The peak position of the \logxp~distributions, \logxpmax, was evaluated.
Figure~\ref{figure:slope} shows
the distribution of \logxpmax~as a function of $Q$
for the HERA data~\cite{H1frag,Zcoh,H1coh} 
and of $\sqrt{s}$ for the $e^+e^-$ data.
Over the range shown
the peak moves from $\simeq$~1.5 to 3.0, equivalent to
the position of the maximum of the
corresponding momentum spectrum increasing from
$\simeq$~400 to 900~MeV.
The HERA data points are
consistent with those from TASSO~\cite{TASSO}
data and a clear agreement in the rate of
growth of the HERA points with
the $e^+e^-$ data~\cite{TASSO,hump} is observed.

The increase of \logxpmax~can be approximated phenomenologically
by the straight line fit
$ \logxpmax\;=\; b\: \ln(Q)+c $
also shown in figure~\ref{figure:slope}.
Also shown is the statistical fit to the data
when $b=1$  which would be the case if the QCD
cascade
was of an incoherent nature, dominated by cylindrical phase space. (A
discussion of phase space  effects is given in~\cite{boudinov}.)
In such a case, the logarithmic particle momentum spectrum would be
peaked at a constant value of momentum, independent of $Q$.
The observed gradient is clearly inconsistent with $b=1$
and therefore inconsistent with cylindrical phase space thus
supporting the coherent nature of gluon radiation.
\section{BFKL versus DGLAP Evolution}
In the DGLAP parton evolution scheme~\cite{dglap}
the parton cascade follows a strong ordering in transverse
momentum
$k_{Tn}^2 \gg k_{Tn-1}^2 \gg... \gg k_{T1}^2$,
while there is only a soft
(kinematical) ordering for the fractional momentum
$x_n<x_{n-1}<...<x_1$.
However for low-$x$ at HERA the BFKL scheme~\cite{bfkl} 
could well be the dominant scheme.
In this scheme the cascade follows a strong ordering in fractional
momentum
$x_n \ll x_{n-1} \ll... \ll x_1$,
while there is no ordering in transverse
momentum.

BFKL evolution can be enhanced by
studying
DIS events which contain an identified jet of longitudinal momentum
fraction $x_{jet}=p_z(j)/E_{proton}$ (in the proton direction)
which is large compared to Bjorken $x$~\cite{mueller}.
By tagging a forward jet
with $p_T(j)\simeq Q$ this allows minimal phase space for
DGLAP evolution while the condition $x_{jet}\gg x$ leaves BFKL evolution
active. This leads to the forward
jet production cross section in BFKL dynamics being larger than that of
the ${\cal O}(\alpha_S^2)$ QCD calculation with DGLAP evolution~\cite{MZ-prl}.

\begin{figure}[t]
\begin{center}
\hspace*{0in}
\centerline{\epsfig{file=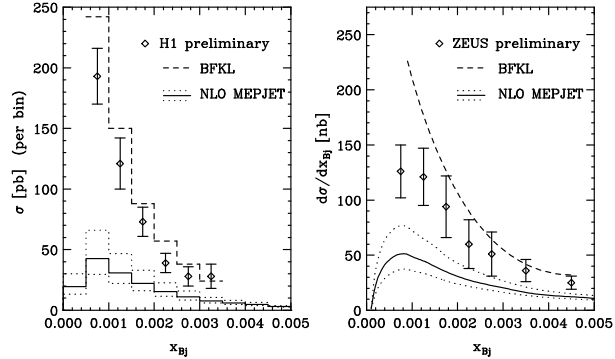,width=8cm}}
\caption{\it
Forward jet cross section at HERA as a function of Bjorken $x$ within
(a)
the H1~\protect\cite{H1-forward} and (b) the ZEUS~\protect\cite{woelfle}
acceptance cuts.
The BFKL result of Bartels et al.~\protect\cite{bartelsH1}
is shown as
the dashed line.
The solid and dotted line give the NLO MEPJET result and a measure for
the uncertainity of NLO prediction through changes in the choice of
scale.
\label{fig:h1comp}
}
\end{center}
\end{figure}

In Fig.~\ref{fig:h1comp}, recent data from H1~\cite{H1-forward} and
ZEUS~\cite{woelfle} are compared with BFKL predictions~\cite{bartelsH1}
and fixed order QCD predictions as calculated with the
MEPJET~\cite{mepjet}
program at NLO. The conditions $p_T(j)\simeq Q$ and
$x_{jet}\gg x$ are satisfied in the two experiments by slightly
different
selection cuts. H1 selects events with a forward jet of $p_T(j)>3.5$~GeV
(in the angular region $7^o < \theta(j) < 20^o$) with
\begin{equation}
   0.5  <  p_T(j)^2/Q^2\; < \; 2\;, \qquad \qquad
   x_{jet}  \simeq  E_{jet}/E_{proton} > 0.035\;; \label{eq:fj-H1}
\end{equation}
while ZEUS triggers on somewhat harder jets of $p_T(j)>5$~GeV
and $\eta(j)<2.4$ with
\begin{equation}
   0.5  <  p_T(j)^2/Q^2\; < \; 4\;, \qquad \qquad
   x_{jet}  =  p_z(j)/E_{proton} > 0.035\;. \label{eq:fj-ZEUS}
\end{equation}

Fig.~\ref{fig:h1comp} shows that 
both experiments observe substantially more forward jet events
than expected from NLO QCD. A very rough estimate of the uncertainty of
the NLO calculation is
provided by the two dotted lines, which correspond to variations
by a factor 10 of the renormalisation and factorisation scales
$\mu_R^2$ and $\mu_F^2$.
A recent BFKL calculation (dashed lines) gives a better agreement
with the data.
The overall normalisation in this calculation
is uncertain and the agreement may be
fortuitous, indeed it should also be noted
that both experiments observe more centrally
produced dijet events than predicted by the NLO QCD calculations.
Further careful investigation is necessary before claiming that BFKL is the
mechanism for this enhanced forward jet production.

\section{Conclusions}
To understand the underlying QCD processes in DIS
 it is necessary to study the
hadronic final state.
At the current level of understanding, QCD works well and 
describes the HERA data. 
As the precision of the HERA data improves and the NLO QCD calculations
become available the framework of QCD is being tested more thoroughly. 
As yet it is not possible to say conclusively whether the effects of
BFKL dynamics are being observed in the HERA data, much theoretical and
experimental work is in progress to define and measure
variables that will allow a definitive statement.
\vskip 2.cm \leftline{\Large\bf Acknowledgements} \vskip4.mm \noindent
I would like to thank the organising committee for inviting me to the
school and providing such a pleasant and enjoyable environment.


\begin{thebibliography}{99}
\parskip=0pt
{\footnotesize
\bibitem{b:VXD} C. Alvisi et al., Nucl. Inst. Meth. A305 (1991) 30.

\bibitem{b:CTD} N. Harnew et al., Nucl. Inst. Meth. A279 (1989) 290;
                B. Foster et al., Nucl.~Phys.~B (Proc.~Suppl.) 32 (1993) 181;
                B. Foster et al., Nucl.~Inst.~Meth. A338 (1994) 254.

\bibitem{b:CAL} M. Derrick et al., Nucl. Inst. Meth. A309 (1991) 77;
                A. Andresen et al., Nucl. Inst. Meth. A309 (1991) 101;
                A. Bernstein et al., Nucl. Inst. Meth. A336 (1993) 23. 


\bibitem{z_shift}
ZEUS Collab., M.\ Derrick et al., Z.~Phys.{ C72} (1996) 399.

\bibitem{b:LUMI} J. Andruszk\'{o}w et al., DESY 92-066 (1992).

\bibitem{b:Detector} ZEUS Collab., The ZEUS Detector, 
                     Status Report 1993, DESY 1993.

\bibitem{H1app}
H1 Collaboration, I.\ Abt et al., DESY preprint 93-103 (1993), 
\NIM{386}{310}{97} (Vol 1) and {\it ibid.} p.348 (Vol 2).

\bibitem{DA} S.~Bentvelsen, J.~Engelen and P.~Kooijman, 
Proceedings of the 1991 Workshop on Physics at HERA, DESY Vol.~1~(1992)~23. 

\bibitem{dglap}
G. Altarelli and G. Parisi, Nucl. Phys. { 126} (1977) 297;
V.N. Gribov and  L.N. Lipatov, 
Sov. J. Nucl. Phys. { 15} (1972) 438 and 675;
Yu. L. Dokshitzer, Sov. Phys. { JETP 46} (1977) 641.

\bibitem{bfkl}
E.A. Kuraev, L.N. Lipatov and V.S. Fadin, 
Sov. Phys. { JETP 45} (1977) 199;
Y.Y. Balitsky and L.N. Lipatov, 
Sov. J. Nucl. Phys. { 28} (1978) 282.

\bibitem{JADE}
JADE Collab., W.~Bartel et al., Z.~Phys.~C33 (1986) 23;
\newline JADE Collab., S.~Bethke et al., Phys.~Lett.~B213 (1988) 235.

\bibitem{PROJET}
{D.~Graudenz}{,}
 CERN-TH.7420/94 (1994), to appear in Comp.~Phys.~Comm.

\bibitem{DISJET}
{T.~Brodkorb and E.~Mirkes}{,}
 Univ. of Wisconsin, MAD/PH/821 (1994).

\bibitem{h1alphas} H1 Collab., T. Ahmed et al.,
                   Phys. Lett. B346 (1995) 415.
%
\bibitem{zeusalphas}
{\rm ZEUS} Collab., M. Derrick et al.,
 Phys. Lett. B363 (1995) 201.

\bibitem{pdg} Particle Data Group, R.M. Barnett et al.,
 Phys. Rev. D54 (1996).

\bibitem{mepjet} E. Mirkes and D. Zeppenfeld, 
Phys. Lett. B380 (1996) 205. 
%
\bibitem{disent} S.~Catani and M.~Seymour,
                 Nucl. Phys. B 485 (1997) 291 and hep-ph/9605323;
%
\bibitem{ktalgo} S. Catani, Y. Dokshitzer and B. Webber,
                  Phys. Lett. B285 (1992) 291.
%
\bibitem{durweb} B.~R.~Webber, J. Phys. G19 (1993) 1567.
%
\bibitem{trefzger}  
T. Trefzger, 
Proc. of the Int. Workshop on DIS and Related Phenomena,
ed. G. D'Agostini and A. Nigro, Rome, 1996.

\bibitem{feyn} R.P. Feynman, ``Photon-Hadron Interactions'', Benjamin, N.Y.
(1972).

\bibitem{bethke} S.~Bethke, Proceedings {\em QCD~94}, Montpellier,
                 ed. S~Narison, Nucl. Phys. B (Proc. Suppl.) 39~B, C~(1995),
                 p. 198.

\bibitem{H1shapes} H1 collab., C.~Aloff et al., DESY-97-098.
[hep-ex/9706002]

\bibitem{webber} B.R.~Webber, Proceedings  
                 {\em Workshop on Deep Inelastic Scattering
                 and QCD}, Paris (1995), eds. J.F.~Laporte and Y.~Sirois,
                 p. 115; 
                  M.~Dasgupta and B.R.~Webber, 
                 preprint Cavendish-HEP-96/5 and hep-ph/9704297.

\bibitem{dokshitzer} Yu.L.~Dokshitzer and B.R.~Webber, 
                     Phys. Lett. B 352 (1995) 451;
                     Z.~Kunszt, P.~Nason, G.~Marchesini and B.R.~Webber,
                     {\em $Z$ Physics at LEP~1},
                     eds. G.~Altarelli, R.~Kleiss and C.~Verzegnassi,
                     CERN~89-08, vol.~1, p.~373.

\bibitem{eedata_shape} PLUTO Collaboration, Ch.~Berger et al.,
                 Z. Phys. C 12 (1982) 297;
                 Mark~II Collaboration, A.~Peterson et al.,
                 Phys. Rev. D 37 (1988) 1;
                 TASSO Collaboration, W.~Braunschweig et al.,
                 Z. Phys. C 45 (1989) 11 and
                 Z. Phys. C 47 (1990) 187; 
                 AMY Collaboration, Y.K.~Li et al.,
                 Phys. Rev. D 41 (1990) 2675;
                 DELPHI Collaboration, P.~Abreu et al.,
                 Z. Phys. C 73 (1997) 229.

\bibitem{basics} Yu.~Dokshitzer, V. Khoze, A. Mueller and S. Troyan, 
``Basics of Perturbative QCD'', Editions Fronti\`{e}res, Gif-sur-Yvette, 
France (1991).
\bibitem{webnas} G.~Altarelli et al., Nucl. Phys. B160 (1979) 301;
P.~Nason and B.~R.~Webber, Nucl. Phys. { B421}~(1994) 473.
\bibitem{eedis} Yu.~Dokshitzer et al., Rev. Mod. Phys. 60 (1988) 373.
\bibitem{anis} A. V. Anisovich et al., Il Nuovo Cimento, A106 (1993) 547.
\bibitem{char} K.\ Charchu{\l}a, J. Phys. G19 (1993) 1587.
\bibitem{H1frag} H1 collab., C.~Adloff et al., DESY-97-158
[hep-ex/9707005 fig.~2 amended 15th August 1997]
\bibitem{Zfrag} ZEUS collab., J.~Breitweg et al., DESY-97-183.
\bibitem{Dirk} D.~Graudenz, CERN--TH/96--52;
D.~Graudenz, CYCLOPS program and private communication.

\bibitem{binnewies} J.~Binnewies et al., Z Phys C65 (1995) 471.
\bibitem{Zcoh}
        ZEUS Collaboration, M.\ Derrick et al., \ZP{67}{93}{95}.
\bibitem{H1coh}
H1 Collaboration, S.\ Aid et al., \NP{445}{3}{95}.

\bibitem{TASSO} TASSO Collab., W.~Braunschweig et al., Z. Phys. C47
(1990) 187;
TASSO Collab., W.~Braunschweig et al., Z. Phys. C22 (1984) 307.

\bibitem{hump} OPAL Collab., M.~Akrway et al., Phys. Lett. B247 (1990)
617. TOPAZ Collab., R.~Itoh et al., Phys. Lett. B345 (1995) 335.

\bibitem{boudinov} E.R. Boudinov, P.V. Chliapnikov and V.A. Uvarov, 
Phys. Lett. B309 (1993) 210. 

\bibitem{mueller}
A.H. Mueller, Nucl. Phys. B (Proc. Suppl.) { 18C} (1990) 125;
J. Phys. { G17} (1991) 1443;
J. Kwiecinski, A.D. Martin and P.J. Sutton, Phys. Rev. { D46}
(1992) 921;
W.K. Tang, Phys. Lett. { B278} (1992) 363. 


\bibitem{MZ-prl}
E.~Mirkes and D.~Zeppenfeld, Phys.~Rev.~Lett.~{ 78} (1997) 428 
[hep-ph/9609231].

\bibitem{H1-forward}
M.~Wobisch, to appear in proceedings of DIS97 (and references therein).

\bibitem{woelfle}
S. W\"olfle,  to appear in proceedings of DIS97

\bibitem{bartelsH1}
J.~Bartels et al., Phys. Lett. { B384} (1996) 300 [hep-ph/9604272]. 
}
\end{thebibliography}
\end{document}